\documentclass{aa}
\usepackage{amssymb}
\usepackage{amsmath}
\usepackage{graphicx}
\usepackage{txfonts}
\usepackage{natbib}

\begin{document}

\title{The \textsl{XMM-Newton} view of the Crab}

\author{M.G.F.\ Kirsch\inst{1},
  G.\ Sch\"onherr\inst{1,2}, E.\ Kendziorra\inst{2},
  M.J.\ Freyberg\inst{3}, M.\ Martin\inst{2}, 
  J.\ Wilms\inst{4}, K.\ Mukerjee\inst{5}, M.G.Breitfellner\inst{1}, M.J.S.\
  Smith\inst{1} and R.\ Staubert\inst{2} }

\offprints{M.G.F.\ Kirsch, mkirsch@sciops.esa.int}

\institute{European Space Astronomy Centre (ESAC), ESA, Apartado -- P.O. Box 50727, 28080 Madrid, Spain
\and
  Institut f\"ur Astronomie und Astrophysik der Universit\"at T\"ubingen,
  Abteilung Astronomie, Sand 1, 72076 T\"ubingen, Germany
\and
  Max-Planck-Institut f\"ur extraterrestrische Physik,
  Giessenbachstrasse, 85748 Garching, Germany
\and
  Department of Physics, University of Warwick, CV4 7AL, UK
\and
  Department of Astronomy and Astrophysics, Tata Institute of
  Fundamental Research, Colaba, Mumbai-400005, India }

   \date{Received 28.12.2005; accepted 15.03.2006}

  \abstract
   {}
   {We discuss the current X-ray view of the Crab Nebula and Pulsar,
     summarising our analysis of observations of the source with the
     EPIC-pn camera on board the \textsl{XMM-Newton} observatory.
     Different modes of EPIC-pn were combined in order to yield a
     complete scenario of the spectral properties of the Crab resolved
     in space and time (pulse phase). In addition we give a description 
     of the special EPIC-pn Burst mode and guidance for data reduction in 
     that mode.}
   {We analysed spectra for the nebula and pulsar separately in the
     0.6--12.0\,keV energy band. All data were processed with the
     \textsl{SAS 6.0.0} \textsl{XMM-Newton} Scientific Analysis System
     package; models were fitted to the data with \textsl{XSPEC} 11.
     The high time resolution of EPIC-pn in its Burst mode (7\,$\mu$s)
     was used for a phase resolved analysis of the pulsar spectrum,
     after determination of the period with epoch folding techniques.
     Data from the Small Window mode were processed and corrected for pile-up
     allowing for spectroscopy simultaneously resolved in space and time.
     }
   {The spatial variation of the spectrum over the entire region of
     the Crab shows a gradual spectral softening from the inner pulsar
     region to the outer nebula region with a variation in photon
     index, $\Gamma$, from 2.0 to 2.4. Pulse phase resolved spectroscopy
     of the Crab Pulsar reveals a phase dependent modulation of the
     photon index in form of a significant hardening of the spectrum
     in the inter-peak phase from $\Gamma =1.7$ during the pulse peak
     to $\Gamma =1.5$.}
   {}

\keywords{stars: neutron stars -- pulsars: individual: PSR 0531+21 -- supernova
  remnants: Crab -- X-rays:stars, -- instruments: EPIC-pn, -- data analysis: Burst mode
}

\authorrunning {M.G.F.\ Kirsch et al.}
\titlerunning  {The \textsl{XMM-Newton} view of the Crab}
\maketitle
%

\section{Introduction}

Since the discovery of the Crab Pulsar in 1968 \citep{Staelin
  Reifenstein}, the Crab has been one of the best studied objects in
the sky and it remains one of the brightest X-ray sources regularly
observed.  With an X-ray luminosity of
$\sim$$5\,10^{37}\,\text{erg}\,\text{s}^{-1}$, the Crab emits a
significant fraction of its energy output in the X-ray band (the total
luminosity of the Crab is
$\sim$$2\,10^{38}\,\text{erg}\,\text{s}^{-1}$). As a standard candle
for instrument calibration, it has been repeatedly studied by many
astronomy missions. The 33\,ms Crab Pulsar has been observed in almost
every band of the electromagnetic spectrum. Its pulse profile exhibits
a double peaked structure with a phase separation of 0.4 between the
first and the second pulse.

Pulse phase resolved X-ray spectroscopy of the Crab was performed for
the first time with OSO 8 \citep{Pravdo Serlemitsos 81}. Analyses with
a finer phase resolution were published from \textsl{RXTE}
\citep{1997ApJ...491..808P} and from \textsl{Beppo-SAX}
\citep{2000A&A...361..695M}. Both measurements showed a significant
hardening of the spectrum in the inter-peak region.
\textsl{RXTE} and \textsl{Beppo-SAX} both were unable to resolve the
Crab spatially.  Recent \textsl{Chandra} observations have provided
spatially resolved spectroscopy of the Crab
\citep{2004ApJ...609..186M}.  These observations, however, suffered
from pile-up which had to be especially corrected for. Similar correction
methods are discussed for the case of the EPIC-pn Small Window mode in
Sect.~\ref{sec:spatialsw}.

With the European Photon Imaging Camera
\citep[EPIC;][]{strueder01,turner01} in its Burst mode,
\textsl{XMM-Newton} provides the unique possibility to measure the
Crab spectrum without pile-up when combining one dimensional spatial
resolution with a high time resolution ($7\,\mu$s), allowing for
spectroscopy simultaneously resolved in space and time (phase). In addition,
we derived two--dimensionally resolved spectra using bootstrapping
techniques and Monte Carlo simulations to correct \textsl{XMM-Newton}
EPIC-pn Small Window mode data.

This paper is organised as follows. In Sect.~\ref{sec:obs} we give a
description of our observations and the \textsl{XMM-Newton}
instruments, followed by some technical comments on data analysis of
Burst mode observations in Sect.~\ref{sec:data}.
Our results from timing and spectral
analysis of the Crab are presented in Sect.~\ref{sec:results}.

\section{Observations performed by \textsl{XMM-Newton}}\label{sec:obs}

\textsl{XMM-Newton} \citep{jansen01} was launched in 1999 December
with an Ariane 5 rocket from French Guyana. It operates six
instruments in parallel on its 48\,hour highly elliptical orbit:
three Wolter type~1 telescopes, with 58 nested mirror shells each,
focus X-ray photons onto the five X-ray instruments of the EPIC
\citep{strueder01,turner01} and the Reflecting Grating Spectrometers
\citep[RGS;][]{herder01}.  In addition, a 30\,cm Ritchey Chr\'etien
optical telescope, the Optical Monitor, is used for optical observations
 \citep[OM;][]{mason01}.  EPIC consists of three
cameras: The two EPIC-MOS cameras use Metal-Oxide Semiconductor CCDs
as X-ray detectors, while the EPIC-pn camera is equipped with a
pn-CCD. Both have been especially developed for \textsl{XMM-Newton}
\citep{pf99,me99,turner01}.

EPIC provides spatially resolved spectroscopy over a field-of-view of
$30'$ with moderate energy resolution.  The EPIC cameras can be
operated in different observational modes related to different readout
procedures.  Detailed descriptions of the various readout modes of
EPIC-pn and their limitations are given by \citet{kendziorra 99},
\citet{kuster 99} and \citet{ehle03}.  The EPIC-pn camera, which
provides the highest time resolution (Timing mode: 30\,$\mu$s, Burst
mode: 7\,$\mu$s) and moderate energy resolution ($E/\text{d}E =
10$--$50$) in the 0.2--15\,keV energy band is the ideal instrument to
observe the Crab. In Burst mode the pile-up limit for a point source
is $60000\,\text{counts}\,\text{s}^{-1}$, corresponding to a maximum
flux of 6.3\,Crab.  The special readout of the Burst mode, however,
leads to a loss of spatial resolution in the shift direction.
Moreover, the lifetime in Burst mode is only 3\%, limiting the mode to
observations of very bright sources such as the Crab or very bright
transients.

The Crab was observed with the EPIC-pn camera in Burst mode during
\textsl{XMM-Newton}'s revolutions 234 and 411.  Observations were also
performed in the EPIC-pn Timing and Small Window modes, however the
high count rate of the source implies severe pile-up effects which
distort the spectrum and require special treatment in the analysis.
The Timing mode data have not been used here, since they do not
contain additional information compared to the Burst mode data. The
Small Window mode data, however, are used despite of the pile-up
distortion, as they provide a two-dimensional spatial resolution of
the source.

Table~\ref{tab:log} summarises the observations of the Crab with
EPIC-pn which were included in our analysis, listing operational modes
and total exposure times for each individual observation. The exposure
times were corrected for data losses due to telemetry constraints and
detector deadtime.

\begin{table}
\caption{Log of observations of the Crab used in this paper.}\label{tab:log}
\begin{tabular}{lrrcccccc}
\hline
\hline
       \noalign{\smallskip}
   OBS ID  & Rev.    &  Exp. & Mode &Filter & Position \\
          &          & [s] & & & Angle[deg]$^{*}$ \\
            \noalign{\smallskip}
            \hline
            \noalign{\smallskip}
            0122330701 & 56 & 166.5 & SW &T & 269.6 \\
           0135730701 & 234 & 298.1 &  BU &T  & 269.3 \\
           0153750201 & 411 & 105.1 &  BU &M &  267.3 \\
           0153750301 & 411 & 181.7 &  BU &M  & 267.3  \\
           0153750501 & 411 & 201.6 & BU &M   & 267.3   \\
            \noalign{\smallskip}
            \hline

\end{tabular}

BU: Burst, SW: Small Window, T: Thick, M: Medium,
Exp.: Exposure time, taking also the life
time of the different modes into account.

$^{*}$ Rotation between the spacecraft $X$-$Z$-plane and the plane
defined by the spacecraft $X$-axis and celestial North.
\end{table}

\section{Data analysis for EPIC-pn Burst mode}\label{sec:data}
In the following section we describe the special techniques of
data reduction needed for the EPIC-pn Burst mode in order to
be able to derive spectra and pulse profiles of the Crab.

The data sets were processed using the \textsl{XMM-Newton} Scientific
Analysis System, \textit{SAS 6.0.0}.  Event times were corrected to
the solar system barycentre using the SAS tool \textit{barycen}.
Single and double events were considered for the analysis while
$\text{FLAG} = 0$ was chosen to exclude border pixel events, for which
the pattern type is uncertain and thus, the total energy is only known
with lower precision.

\begin{figure}
\resizebox{\hsize}{!}{\includegraphics{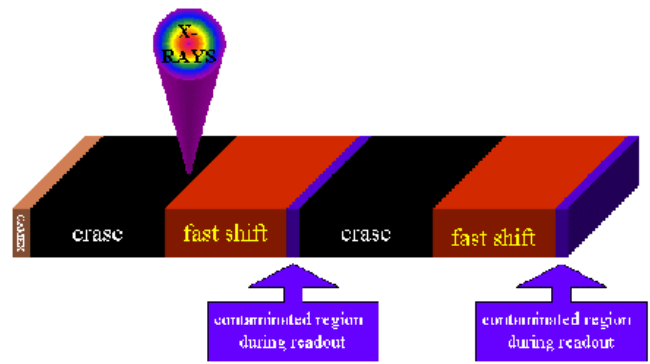}}
\caption{The special readout in Burst mode, which gives the
  possibility to observe very bright sources with negligible pile-up.}
\label{burstmode}
\end{figure}

The Burst mode operates a special readout similar to a tape recorder.
Within 14.4\,$\mu$s, 200 lines are fast-shifted while accumulating
information from the source. Similar to the Timing mode, this leads to
the loss of spatial resolution in the shift direction. The stored
information is then read out in a standard fashion, ignoring the last
20 lines of the CCD because of contamination by source photons
arriving during the readout.  The CCD is then erased by a fast shift
of 200 lines and immediately afterwards the next Burst readout cycle
starts.  Fig.~\ref{burstmode} illustrates this readout procedure.  In
the following we use SAS terminology for the images, i.e., RAWY
represents the CCD line numbers, i.e., the position on the CCD in
shift direction from 1--200 and RAWX represents the CCD column
numbers, i.e., the position perpendicular to the shift direction from
1--64.

Note that events with $\text{RAWY}>180$ cannot be used for any kind of
data analysis, since they contain severely piled-up data as accumulated from
the point-spread function (PSF) wings during the slow readout phase of a Burst cycle. 
(The HEW of \textsl{XMM-Newton's} EPIC-pn PSF is 15$''$ and one EPIC-pn
pixel has a size of $4.128$$\times$$4.128 \square ''$).
These data are therefore not transmitted to ground at all. We
quantified the effect of pile-up in CCD lines $1 \le \text{RAWY} \le 180$ for the measured spectrum of the Crab
by extracting spectra with a sliding box of 13 RAWY pixels length,
shifted in steps of 10 RAWY pixels. An absorbed power law fixing
$N_\mathrm{H}$ at $2.76\,10^{21}\,\text{cm}^{-2}$, as derived from the
nebula region of the Crab (Table~\ref{tab:fit}), was fitted to the
data. The fit results show that the power law index is only affected
for RAWY $>160$ while the normalisation seems to be already
significantly affected as of $\text{RAWY}\sim140$.  Therefore we
recommend for general data analysis in the EPIC-pn Burst mode to
extract spectra only from RAWY-regions $\le 160$ for determination of
the spectral slope and RAWY $\le 140$ for determination of the
normalisation.

\begin{figure}
\centering
\resizebox{\hsize}{!}{\includegraphics [angle=90,bb=75 110 510 750,clip=,]{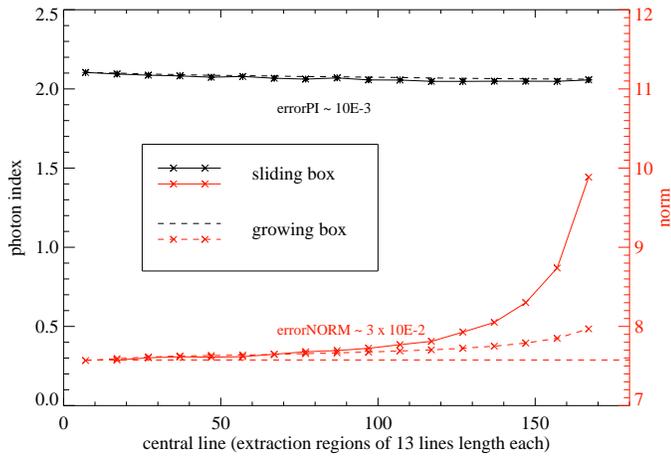}}
\caption{Variation of spectral parameters with extraction region in EPIC-pn
  Burst mode. Spectra have been extracted in a sliding box of 13 RAWY
  pixels in steps of 10 RAWY pixels. An absorbed power law fixing
  $N_\mathrm{H} =2.76\,10^{21}\,\text{cm}^{-2}$ (see
  Table~\ref{tab:fit}) derived from the nebula region of the Crab was
  fitted to the data. In addition we show the variation of the
  spectral parameters extracting spectra from a growing box in order
  to visualise the effect as averaged over columns. Spectra should
  only be extracted for $\text{RAWY} \le 160$ (for determination of
  the photon index) and $\text{RAWY}\le 140$ (for determination of the
  normalisation).  The background was also taken from a sliding box
  that contains in first approximation no source photons. Note that the
  dashed red line without symbols is just to guide the eye with respect
  to the normalisation parameter variation of the growing box (dashed red one with symbols).}
\label{pu_burst}
\end{figure}

The effects due to the special readout in Burst mode on both the
spectral shape and the normalisation of the flux are shown in
Fig~\ref{pu_burst}.
For our spectral analysis we excluded events with $\text{RAWY}\ge142$.

\begin{figure}
\resizebox{\hsize}{!}{\includegraphics{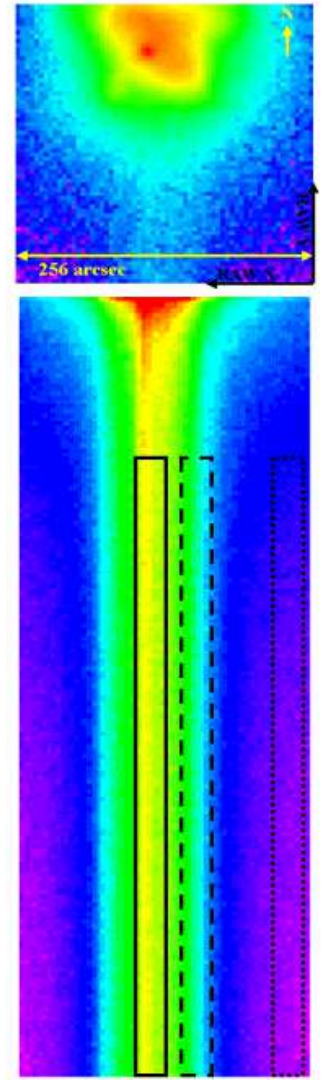}}
\caption{Upper panel: Intensity image of the Crab in EPIC-pn Small Window mode
  (rev.~56). Lower panel: Crab image in Burst mode. All events in the full energy 
range (0.2--15\,keV) were
  used. The solid box indicates
  the pulsar extraction region, while the dashed and dotted boxes are
  used for nebula and background data extraction respectively. }
\label{ori}
\end{figure}

The Burst mode in EPIC-pn provides spatial resolution only
perpendicular to the CCD readout direction, i.e., in the RAWX
direction, which depends on the spacecraft's position angle with
respect to the source. To visualise the orientation of the Crab Nebula
with respect to the readout direction, we compare in Fig.~\ref{ori} a
Burst mode image of the Crab (revolution~234) to an image obtained in
Small Window mode (revolution~56) which was observed under nearly the
same position angle (see Table~\ref{tab:log}). The additional
observations in revolution 411 were done with the slight difference of
$2\fdg{}3$ in the position angle. This variation, however, is still
negligible for the purpose of our analysis.  Thus, we can directly
compare the two-dimensional image of the Crab taken in Small Window
mode with the one-dimensional cut acquired during Burst mode
observations.  The extraction regions to derive spectra of the pulsar,
nebula and background are also illustrated in Fig.~\ref{ori}.

\section{Data analysis results}\label{sec:results}

\subsection{Pulse profile and period} 
\begin{figure}
\centering
\resizebox{\hsize}{!}{\includegraphics[angle=90, bb=40 130 510 755, clip=,]{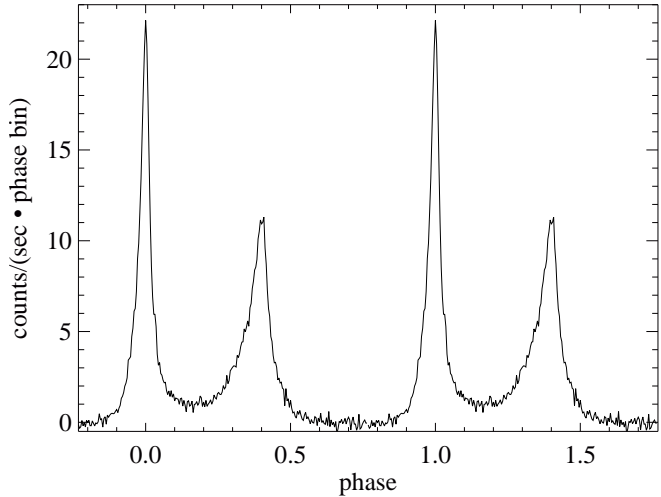}}
\caption{0.2--15\,keV pulse profile of the Crab Pulsar obtained from the three
  observations in rev.~411, using 250 phase bins. The profile is shown
  twice for clarity.}\label{lc}
\end{figure}

We determined the period of the Crab
pulsar using epoch folding techniques. The period \textit{P} was
computed separately for each individual observation in revolutions 234
and 411, using extrapolated values of $\dot{P}$ supplied by the
Jodrell Bank Crab Pulsar Monthly Ephemeris
(\texttt{http://www.jb.man.ac.uk}).  The best pulsar period derived
from the data from revolutions 234 and 411 were 33.52130944(2)\,ms and
33.5341004590(5) ms for the epochs 51988.6405766415 (MJD) and
52340.6825136183 (MJD) respectively. Thus, the relative deviation of
the observed pulse period with respect to the most accurate radio data
value available was $\Delta P/P \lesssim 10^{-9}$.  Data were combined
for detailed analysis from the three observations in revolutions 411.
The folded 0.2--15\,keV pulse profile is shown in Fig.~\ref{lc}.


\subsection{Localisation of the pulsar}
\begin{figure}
\centering
\resizebox{\hsize}{!}{\includegraphics[angle=90, bb=45 130 510 780, ]{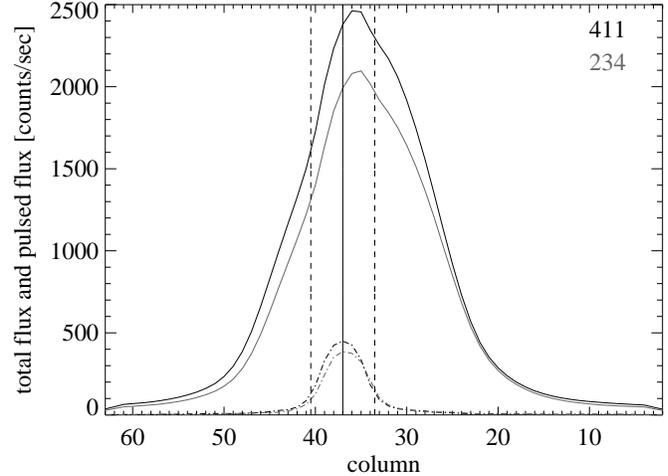}}
\caption{Variation of the 0.5--15\,keV total flux (continuous lines) and the
  pulsed flux (dashed-dotted lines) over the entire Crab. Note that
  the flux units are per $7$ columns window and renormalised to the same extraction area, using a
  normalisation factor of 180/140. For the pulsed flux we subtracted
  the renormalised mean flux from the off-pulse region in phase
  0.568--0.768. The two curves shown are for Rev.~411 and Rev.~234,
  respectively. The difference in flux is caused by the use of different
  filters.  } \label{flux_ccd}
\end{figure}

As outlined above, spatial resolution in the RAWY-direction is lost
when EPIC-pn is operated in Burst mode. To locate the pulsar
without use of the attitude in such an image, events in a running window of 5 columns in
RAWX-direction were folded with the best-fit period. The pulsed flux
for each of those windows was calculated subtracting the off-pulse
region (phase 0.568--0.768) as background contribution. The pulsed
flux reached its maximum, indicating the location of the pulsar, when
the window was centred on column 37. In a next step, the pulsar
spectrum was extracted taking a larger window of seven columns width
($28\farcs{}896$) centred at column 37. The spatial variation of the
total flux and the pulsed flux is shown in Fig.~\ref{flux_ccd}. The
peaks of the distributions of the total flux and the pulsed flux are
not co-aligned, indicating that the pulsar is not located in the centre
of the nebula. The dashed vertical lines show the region chosen for
the extraction of the pulsar spectra.

\subsection{Spectra of the Crab from EPIC-pn in Burst mode}
\label{subsec:spec_burst}
\begin{figure}
\resizebox{\hsize}{!}{\includegraphics [angle=-90, bb=112 45 560 710, clip=,]{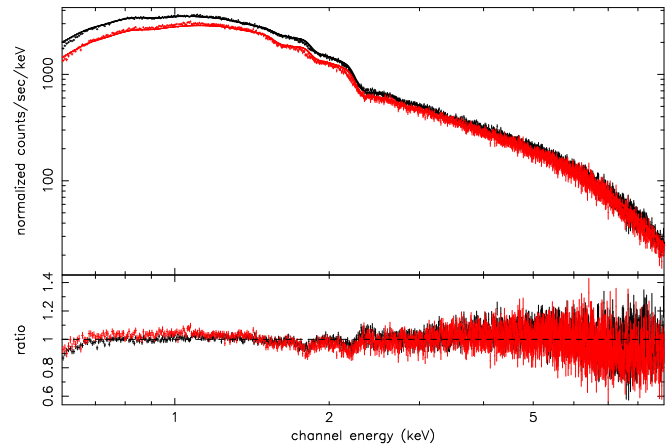}}
\caption{Spectra of the overall Crab region covering
  the 0.6--9.0\,keV energy band. Data points marked in black and red correspond to
  revolutions~411 and~234 respectively. The background contributions
  were subtracted from the data as listed in Table~\ref{tab:fit}.}
\label{all_Crab}
\end{figure}

\begin{figure}
\resizebox{\hsize}{!}{\includegraphics [angle=-90, bb=112 45 560 710, clip=,]{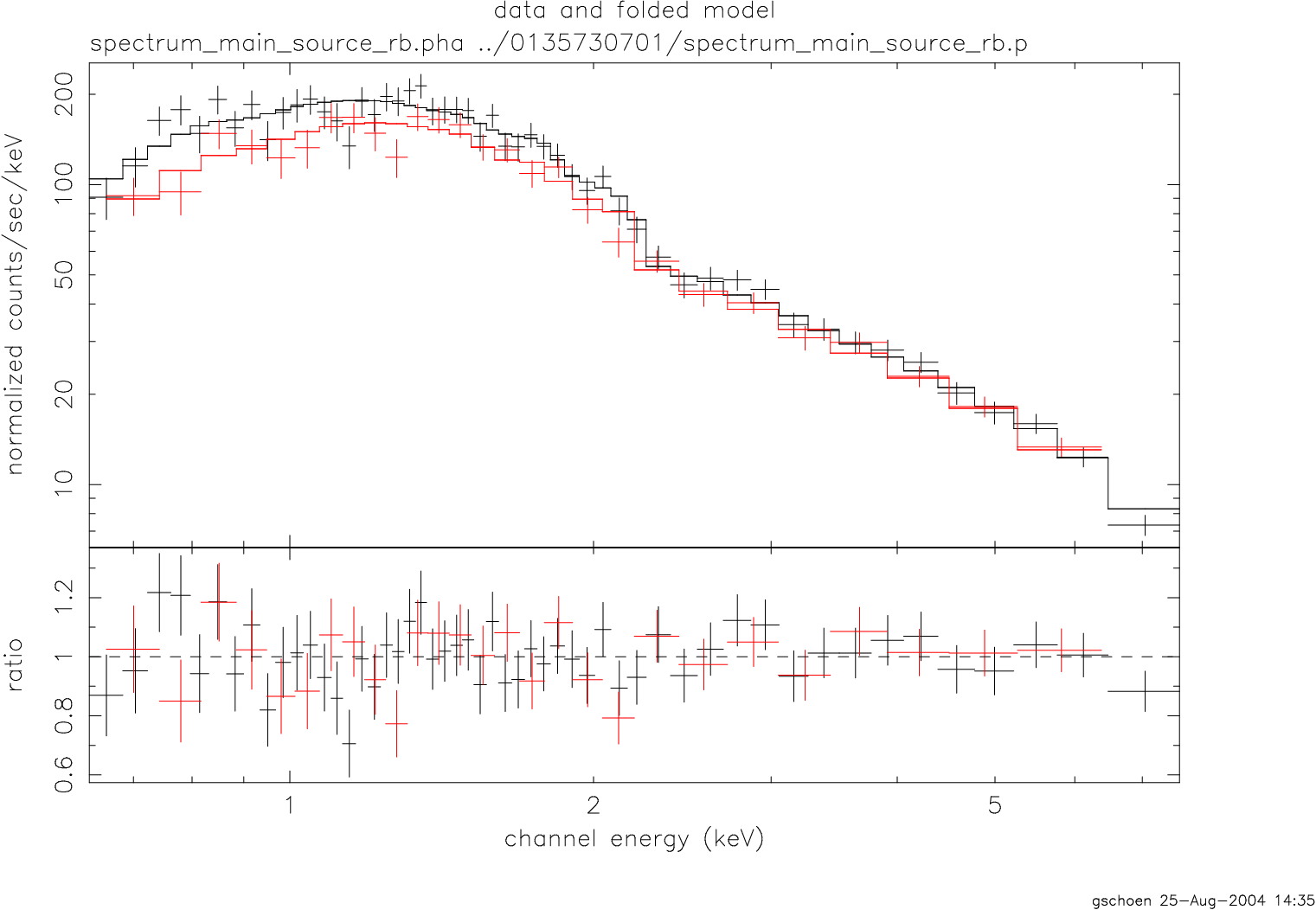}}
\caption{Spectra of the pulsar region covering the
  0.65--8.5\,keV energy band. Data points marked in black and red
  correspond to revolutions 411 and~234, respectively.  The background
  contributions were subtracted from the data as listed in
  Table~\ref{tab:fit}.}
\label{pu_op}
\end{figure}

\begin{figure}
\resizebox{\hsize}{!}{\includegraphics [angle=-90, bb=112 45 560 710, clip=,]{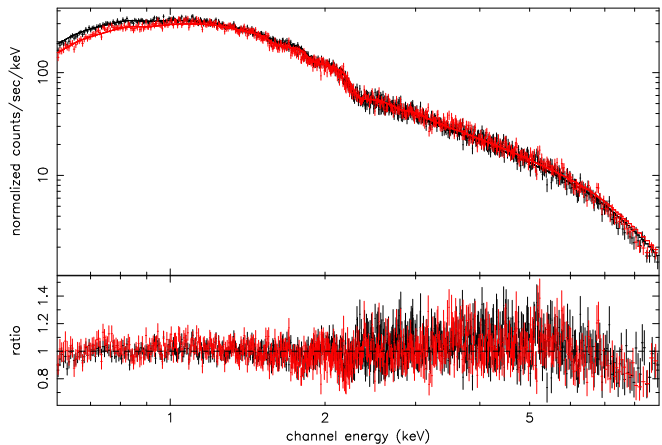}}
\caption{Spectra of the nebula region covering the
  0.6--9.0\,keV energy band. Data points marked in black and red
  correspond to revolutions~411 and~234, respectively. The background
  contributions were subtracted from the data as listed in
  Table~\ref{tab:fit}.}
\label{neb}
\end{figure}

\begin{figure}
\resizebox{\hsize}{!}{\includegraphics [angle=-90, bb=112 45 560 700, clip=,]{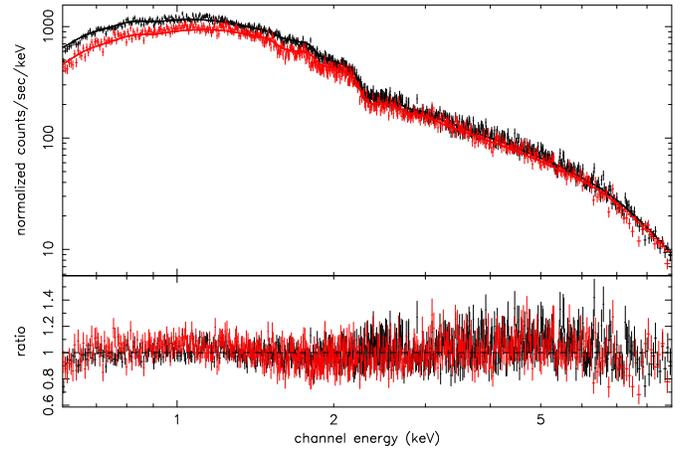} }
\caption{Spectra of the off-pulse region covering
  0.6 -- 9.0\,keV energy band. The data points marked in black and red
  correspond to revolutions~411 and~234, respectively. The background
  contributions were subtracted from the data as listed in
  Table~\ref{tab:fit}.}
\label{op}
\end{figure}

Spectra from different regions of interest of the Crab were extracted
from the single Burst mode observation in revolution 234 and the three
merged observations in revolution 411, as listed in
Table~\ref{tab:fit}.  Note that different ways of background
subtraction were applied (BGSM).  The spectra were fitted with an
absorbed single power law model spectrum in \textsl{XSPEC}. The key
spectral fit parameters are also listed in Table~\ref{tab:fit}. Fits
and model to data ratios are shown in
Figs.~\ref{all_Crab}--\ref{op}.

\subsubsection{Spatial variation of the spectra}
\begin{figure}
\centering
\resizebox{\hsize}{!}{\includegraphics [angle=90, bb=45 70 510 770,clip=,]{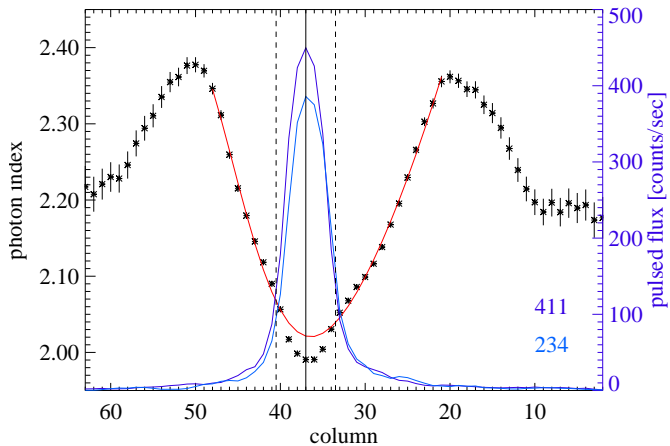}}
\caption{Spatial variation of the photon index (stars) over the Crab
  nebula and pulsed flux (solid blue lines) in the energy range
  0.6--9.0\,keV.  Note that the flux units are renormalised to the
  same extraction area by a factor 180/140. For the pulsed flux we
  subtracted the renormalised mean flux from the off-pulse region
  (phase 0.568--0.768). Note that since the spectral analysis has been performed without
  background subtraction the hardening of the spectral index outside the nebula region
  is influenced by the spectrum of the X-ray background in the Crab direction.}
  \label{CRAB_pi_ccd}
\end{figure}

To study the spatial variations of the source in greater depth,
spectra were extracted from a running window in RAWX-direction of five
columns width.  An absorbed power law model was
fitted to each spectrum, fixing the equivalent hydrogen column density
at $N_\mathrm{H}=2.76\,10^{21}\,\text{cm}^{-2}$ (see
Table~\ref{tab:fit}) as derived from the nebula region of the Crab.
We were able to distinguish four typical regions as (a) background
(columns 1--13 and 55--64) (b) outer nebula (columns 14--22 and
47--54) (c) inner nebula (columns 23--33 and 41--46), and (d) pulsar
(columns 34--40). The variation of the spectral index along those
extraction regions is plotted in Fig.~\ref{CRAB_pi_ccd}.

The spectrum is hardest in the region of maximum flux and pulsed
fraction (see Figs.~\ref{flux_ccd} and~\ref{CRAB_pi_ccd}). The total
variation of the photon index along the Crab is a composed function of
the contributions from pulsar and nebula. While all data points
outside the pulsar region lie on a parabolic curve, the contribution of
the pulsar becomes manifest as an additional dip of the photon index
in columns 34--40. Note that Fig.~\ref{CRAB_pi_ccd} is a result of
spectral analysis without background subtraction in order to show the
hardening of the background outside the nebula region.
\begin{table}
\caption{Extracted spectra from different regions of interest and
  spectral parameters.}\label{tab:fit}
\begin{tabular}{lccccccc}
\hline
\hline
       \noalign{\smallskip}
   Spectrum  &  BGSM$^*$ & $N_\mathrm{H}$  & $\Gamma$ & $\chi$$^{2}_{red}(dof)$ \\
   & & $10^{21}\,\text{cm}^{-2}$$$ &  & \\
            \noalign{\smallskip}
            \hline
            \noalign{\smallskip}
            total Crab & C$-$B & 2.60(1) & 2.046(3) & 1.99(3075)    \\
           pulsar & P$-$O &  2.9(3) & 1.72(5) & 0.91(75) \\
           nebula & N$-$B  & 2.76(4) & 2.263(9) & 1.24(1394)\\
           off-pulse & O$-$B & 2.62(4) & 2.030(9) & 1.23(1296)  \\
            \noalign{\smallskip}
            \hline
  \end{tabular}
\newline $^*$BGSM: Background Subtraction Method: This column
explains which region was used to generate the source and
which for the background spectrum (A-B means: A source; B background
spectrum.)

C: all Crab (columns 22--51), P: pulsar region (columns 34--40), N:
nebula region (columns 22--26), B: background region (columns 3--7),
O: off-pulse region (columns 34--40).

Note that errors are given in terms of 90 \% confidence level.

\end{table}

\subsubsection{Pulse phase spectroscopy}
\begin{figure}
\centering
\resizebox{\hsize}{!}{\includegraphics
  [angle=90, bb=40 80 510 770,clip=,]{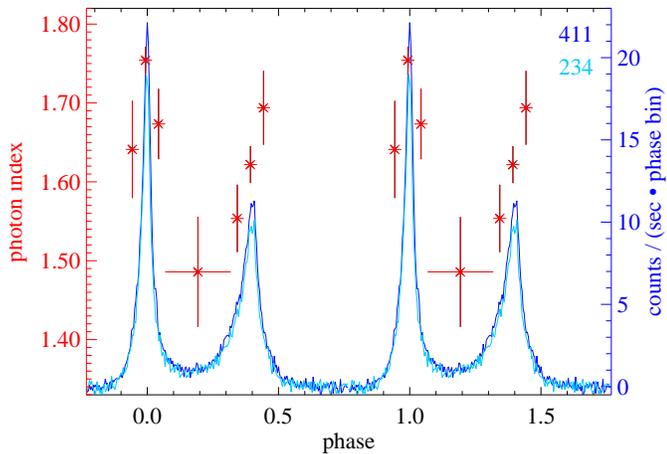}}
\caption{Spectral variation with pulse phase of the Crab Pulsar.
  Photon index (red crosses) in the energy range 0.6--6.5\,keV, pulsed
  flux: blue lines for the two observations in rev 234 and 411. 
  For this analysis only data of the Burst mode
  observations have been used.  The flux has been renormalised to the
  same extraction region by multiplying with a correction factor of
  180/140.  For the pulsed flux we subtracted the renormalised mean
  flux from the off-pulse region with phase 0.568--0.768 and took into
  account that we measure only 77\% of the encirceled energy in the
  extraction region of seven columns.}\label{CRAB_pi_phase_old}
\end{figure}

Events from the pulsar region ($\text{RAWY}<142$, columns 34--40) were
sorted into 7 pulse phase intervals. For each phase interval we
extracted spectra and computed the photon index with $N_\mathrm{H}$
fixed at the value derived from the off-pulse region,
$N_\mathrm{H}=2.62\,10^{21}\,\text{cm}^{-2}$. The reduced $\chi^{2}$
values are generally acceptable ($\chi^2_\text{red}= 1.2$--1.3). For
background subtraction, we used the off-pulse region (columns 34--40
with phase 0.568--0.768).  Plotting the photon index, $\Gamma$,
against the phase intervals, we see a modulation of $\Gamma$ with the
pulse phase (Fig.~\ref{CRAB_pi_phase_old}).  The secondary pulse shows
a harder spectrum than the main pulse. In the intermediate
pulse region the spectrum is found to be harder than in the peak of
the pulses. The spectrum of the intermediate pulse region is harder
than the spectrum of the main pulse by 0.3. The spectrum of the second
pulse is harder than the spectrum of the main pulse by 0.1.

The analysis of \citet{2000A&A...361..695M} did show, that a two
component emission model can account for the variation of the
pulse profile with energy and the behaviour of the spectral index
with phase. However, no complete model for the X-ray emission from
the Crab Pulsar was provided.

\subsection{Spatially resolved spectroscopy of the Small Window mode
  data}
\label{sec:spatialsw}
\begin{figure}
\resizebox{\hsize}{!}{\includegraphics{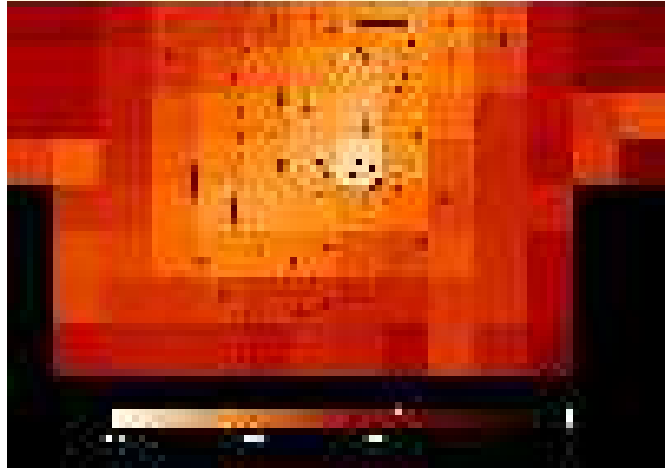}}
\caption{Two dimensional spectroscopy of the Small Window mode data without
  correction for pile-up. One coloured box represents a region of 9 EPIC-pn
  pixels (153.363\,$\square''$), the colour represents the spectral index (see
  colour code scale). The count rate contours in Small Window
  mode are overplotted. } \label{sw_pi_2d}
\end{figure}

In Small Window mode, one EPIC-pn CCD is operated in a window of $64$$\times$$64$ pixels (Fig. \ref{ori}). Events are read out 
after an integration time of $3.9812\,$ms.
Due to the high source intensity, observational data of the Crab taken in Small Window
mode is expected to be severely affected by pile-up (the limiting
count rate in Small Window mode before the onset of pile-up is of the order of $130\,\text{counts}\,\text{s}^{-1}$ for a
point source, \citealt{ehle03}).
However, as outlined above, the Small Window mode provides additional spatial information compared to the Burst mode. 
In analogy to the analysis of Burst mode data (Sect.~\ref{sec:data}, \ref{subsec:spec_burst}), we extracted Small Window mode 
spectra of the Crab (single and double events, FLAG=0) 
and fitted an absorbed power law \textsl{XSPEC} model to the data. The results of our analysis are visualised in a colour 
coded diagram in Fig.~\ref{sw_pi_2d}. The variation of the photon index over the Crab is indicated by the colours 
of the boxes, representing extraction regions of $153.363\,\square''$ each. The observed count rates are overplotted 
as contours. Note that these results are of qualitative nature only and yet have to be corrected for pile-up effects.

\begin{figure}
  \centering \resizebox{\hsize}{!}{\includegraphics[angle=90, bb=45 90
    510 750,clip=,]{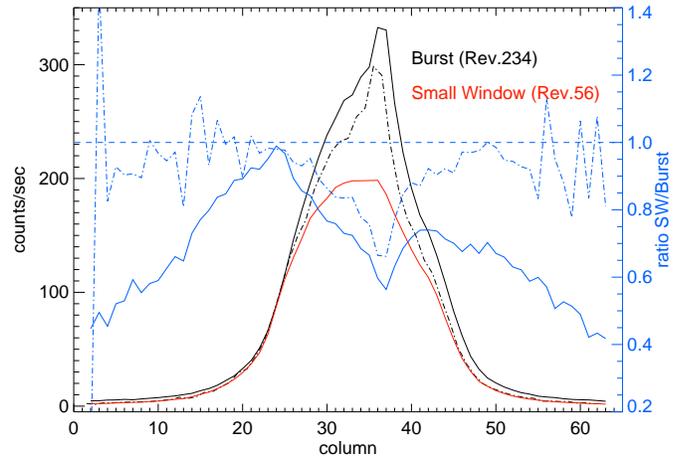}}

\caption{Count rates and ratio of the count rates in Burst and
  Small Window mode, where numbers diverging from one indicate pile-up
  (blue). Black solid: Burst mode count rate, red solid: Small Window mode
  count rate, black dashed: Burst mode count rate corrected (a) for
  pointing difference Burst and Small Window observation of 0.5 pixels between,
  (b) the extraction region has been reduced to RAWY values from 2--11,
  (c) the lower threshold settings of Small Window mode have been also applied to
  the Burst mode data. Blue solid: uncorrected ratio of Burst and Small Window
  mode count rate, blue dashed: corrected ratio of Burst and Small Window mode
  count rate. }\label{pile_up_ratio}
\end{figure}

In order to estimate the level of data contamination due to pile-up we took two different
approaches:
\begin{enumerate}
\item By comparing predicted to measured CCD pattern fractions using the SAS task \textit{epatplot} (Fig.~\ref{pile_up_sw}). 
The change in pattern ratios is
a sensible indicator of the onset of pattern pile-up (Two single photons hitting the detector in the same readout frame 
in adjacent pixels are misinterpreted as a single event of double pattern). 
Since the lower threshold of the EPIC-pn camera
was 80 ADU (400 eV) higher than the default value for the Small Window mode observation
we had to recalibrate \textit{epatplot} using pattern fractions of
non piled-up sources in Small Window mode. This special calibration
will be provided to the public in the near future via the standard
SAS.

\begin{figure}
\resizebox{\hsize}{!}{\includegraphics[angle=90, bb=60 30 580 750,clip=,]{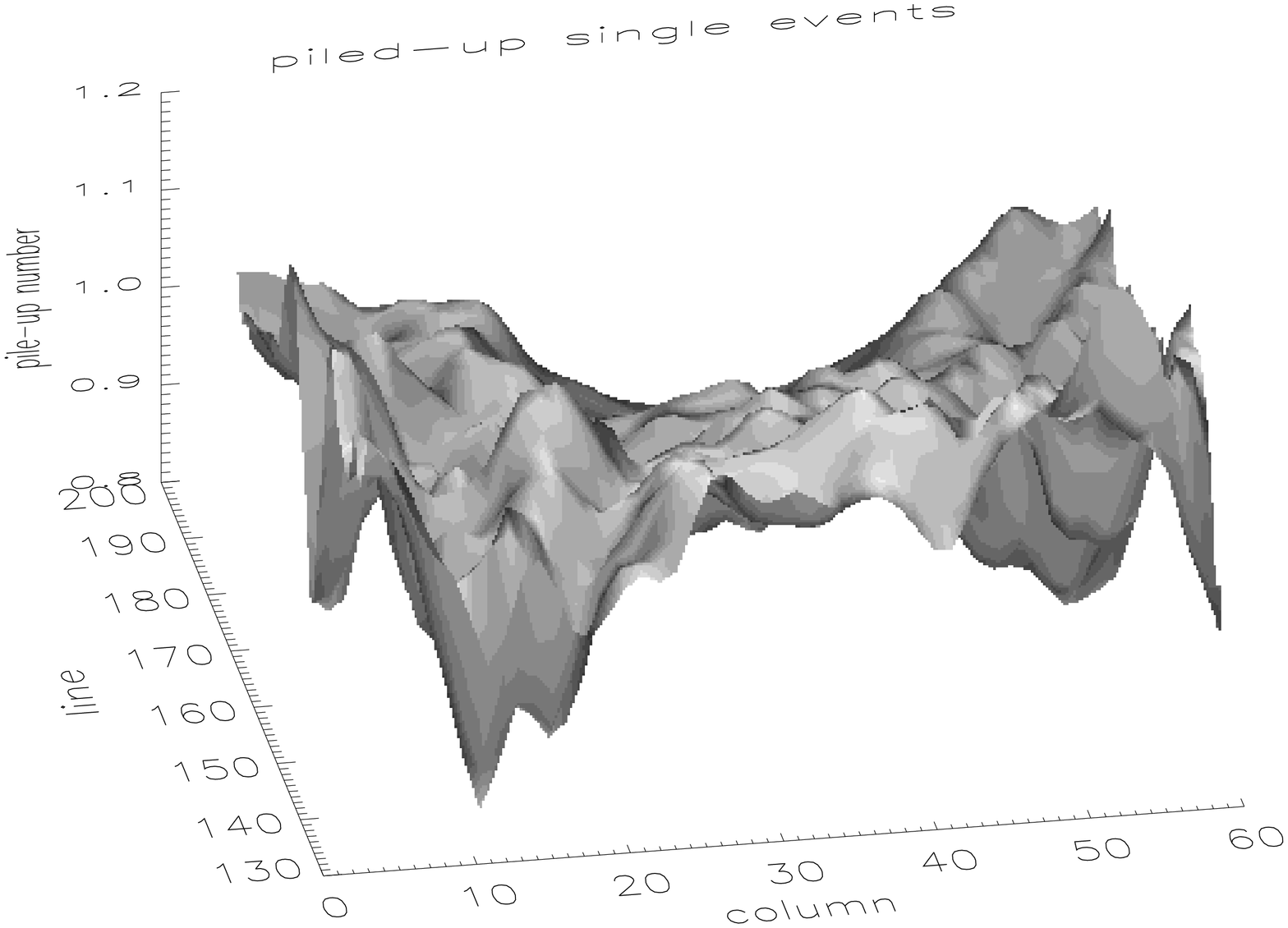}}
\resizebox{\hsize}{!}{\includegraphics[angle=90, bb=60 30 580 750,clip=,]{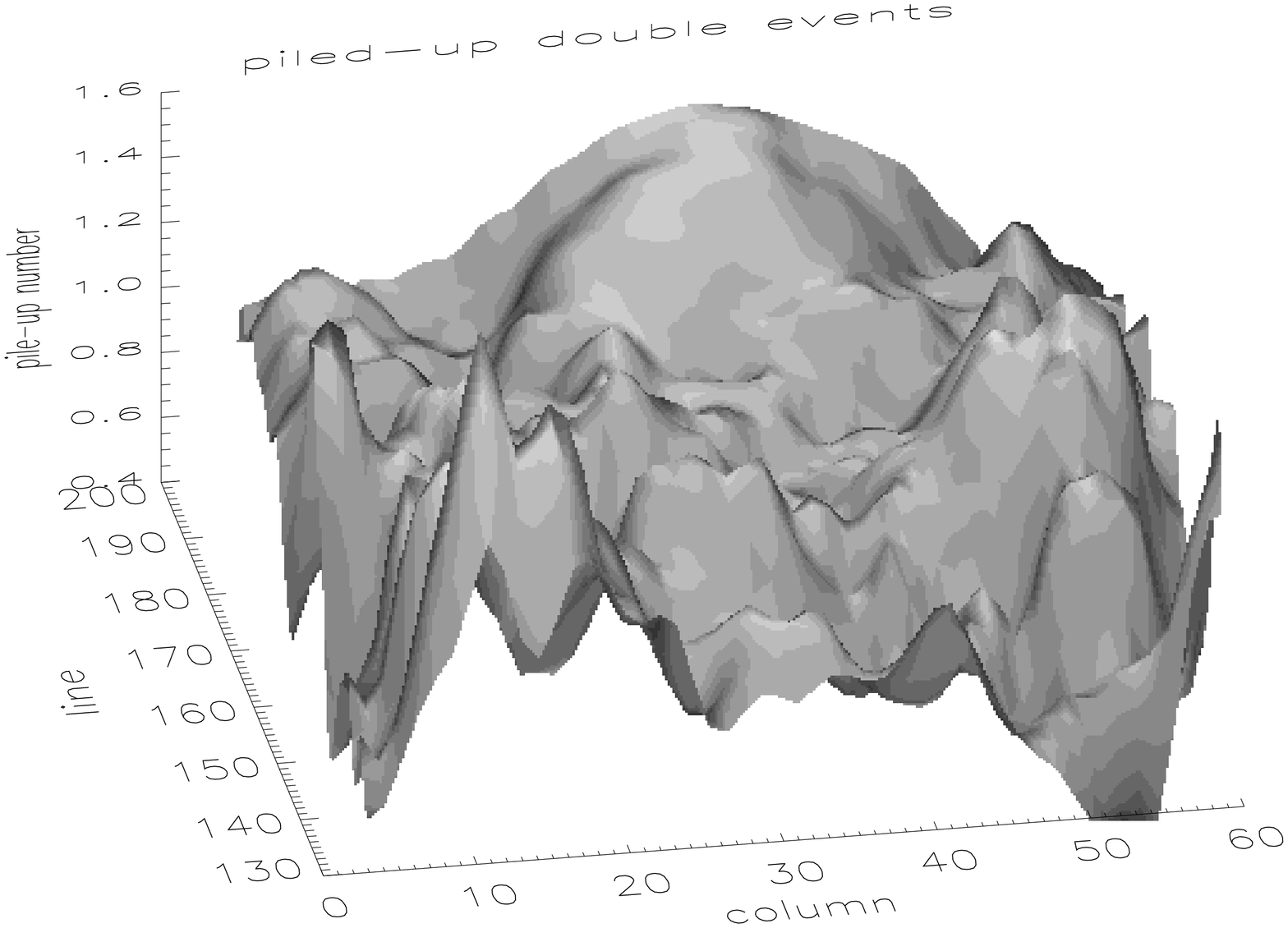}}

\caption{Pile-up assessment of the Crab observations in Small Window mode.
  We compared the predicted with the measured pattern fractions using the
  \textit{SAS}-task \textit{epatplot}. \textit{epatplot} was
  operated on a grid of event regions with $7\times7$ pixels
  ($834.979\,\square''$).  The upper panel shows the deficit in single
  events, the lower the increase of double events. Both is an
  indication for pile-up.}
  \label{pile_up_sw}
\end{figure}

\item Analysing the loss of events due to pile-up by determining the
  count rate of single columns in Burst mode and comparing them with
  the corresponding columns in Small Window mode (Fig.~\ref{pile_up_ratio}). Note that for this analysis
  only events with $\text{RAWY}\le 11$ have been used for Burst mode
  in order to avoid any effect of pile-up during readout hampering the
  normalisation (Fig.~\ref{pu_burst}).  Fig.~\ref{pile_up_ratio} shows
  the ratio of the count rates in Burst and Small Window mode, where
  numbers diverging from 1.0 indicate pile-up.
\end{enumerate}

Both methods confirm that the complete source area of the Crab observation in
Small Window mode may only be used for analysis when correcting the data for pile-up.
Pile-up effects change both flux and spectral shape of the source data, as with increasing 
count rate the probability increases 
that several low-energy events are wrongly detected as single high-energy photons.

\begin{figure*}
\centering
\includegraphics[bb=80 370 550 698,width=8.9cm,clip=,]{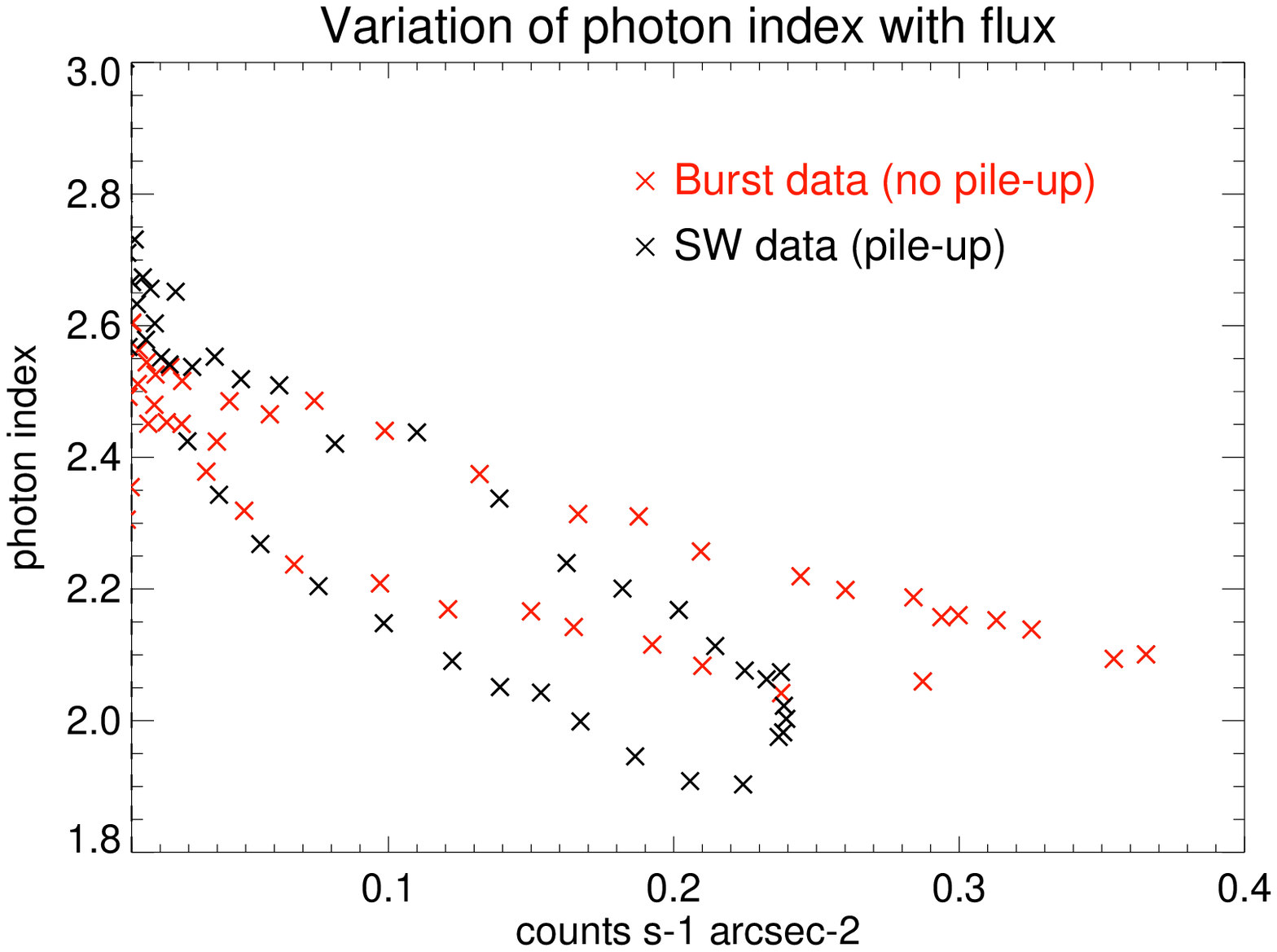}
\includegraphics[bb=80 370 550 698,width=8.9cm,clip=,]{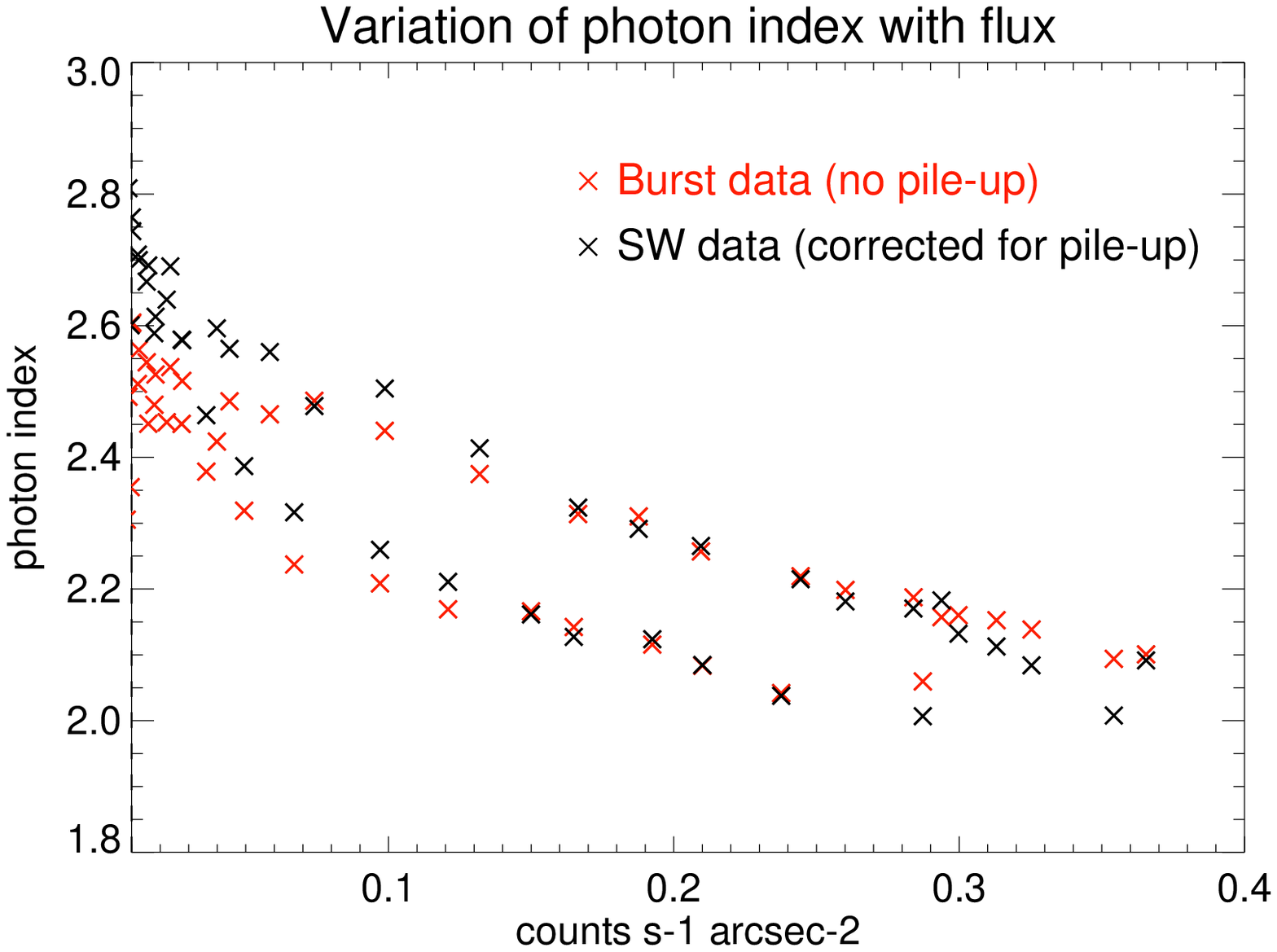}
\caption{Left: Photon indices as a function of count rate 
  for both Small Window and Burst mode. The piled up Small
  Window mode spectra show lower photon indices than the non
  piled-up Burst mode spectra for the same count rates. Right: Same
  data set with pile-up corrections for Small Window data. The Small Window count rates are
  replaced by the correct values from Burst mode, the photon indices
  are corrected according to the function $f(r)$ given by (\ref{eq:corr}).
  Note that the general trend of the data points (lower photon index with higher 
  count rate) reflects the spectral behaviour of the Crab as already shown in 
  Fig.~\ref{CRAB_pi_ccd}.}
 \label{sw_bu_pi}
\end{figure*}

In order to quantify the effects of pile-up on the spectral shape, Burst mode data and the Small
Window mode data have been compared in one spatial dimension, considering spectra extracted from single CCD columns. This approach simulates for the Small
Window mode the loss of spatial resolution along the
shift direction in Burst mode. A direct comparison can thus be made for non piled-up
data from Burst mode and piled-up data from
Small Window mode.  Fig.~\ref{sw_bu_pi} shows at the left the resulting average photon
indices as a function of count rate (column) for both Small Window and
Burst mode data. The piled-up Small Window mode spectra are flatter than the non piled-up Burst mode spectra. This comparison gives a first estimate of the order of pile-up corrections required in general for Small Window mode spectra of Crab-like sources, and will be used in the following to test our theoretical pile-up corrections (Fig.~\ref{sw_bu_pi}, right).

This observational estimate, however, can not be used for a straightforward correction of the 
two-dimensional spectral analysis, where the observed count rate  assumes values up to $1$ 
count$\,$/($\,s\, \square''$) in the central region of the Crab. In a more general approach, 
we performed Monte Carlo simulations, implementing the pile-up
behaviour of the EPIC-pn CCDs. Using non-contaminated Burst mode spectra to fix the input 
spectral shape characterised by various photon indices, and varying the input flux, we simulated 
the effect of the Small Window mode on the source spectra.
Comparing the results from spectral fits of Burst input and
simulated Small Window output spectra, we derived a correction function $f$ for the
photon index with respect to the observed count rate. With the onset of pile-up in Small Window mode, 
the photon index $[p.i.^{SW}(r)]$ decreases linearly with the observed count rate $r$:
\begin{equation}
f(r)=[p.i.^{SW}(r)]\sim[p.i.^{BURST}]-const.-\frac{1}{3}\frac{r}{s\, \square''}
\label{eq:corr}
\end{equation}
The negative constant shift ($const.\sim 0.03$) of the observed with respect to the
true photon index is caused by the unusually high lower threshold of
this Small Window mode observation, which leads to changes
in the fraction of double and single patterns at low energies 
and thus to changes of the spectral shape in the low energy regime.

\begin{figure}
\resizebox{\hsize}{!}{\includegraphics[bb=80 370 550 702,clip=,]{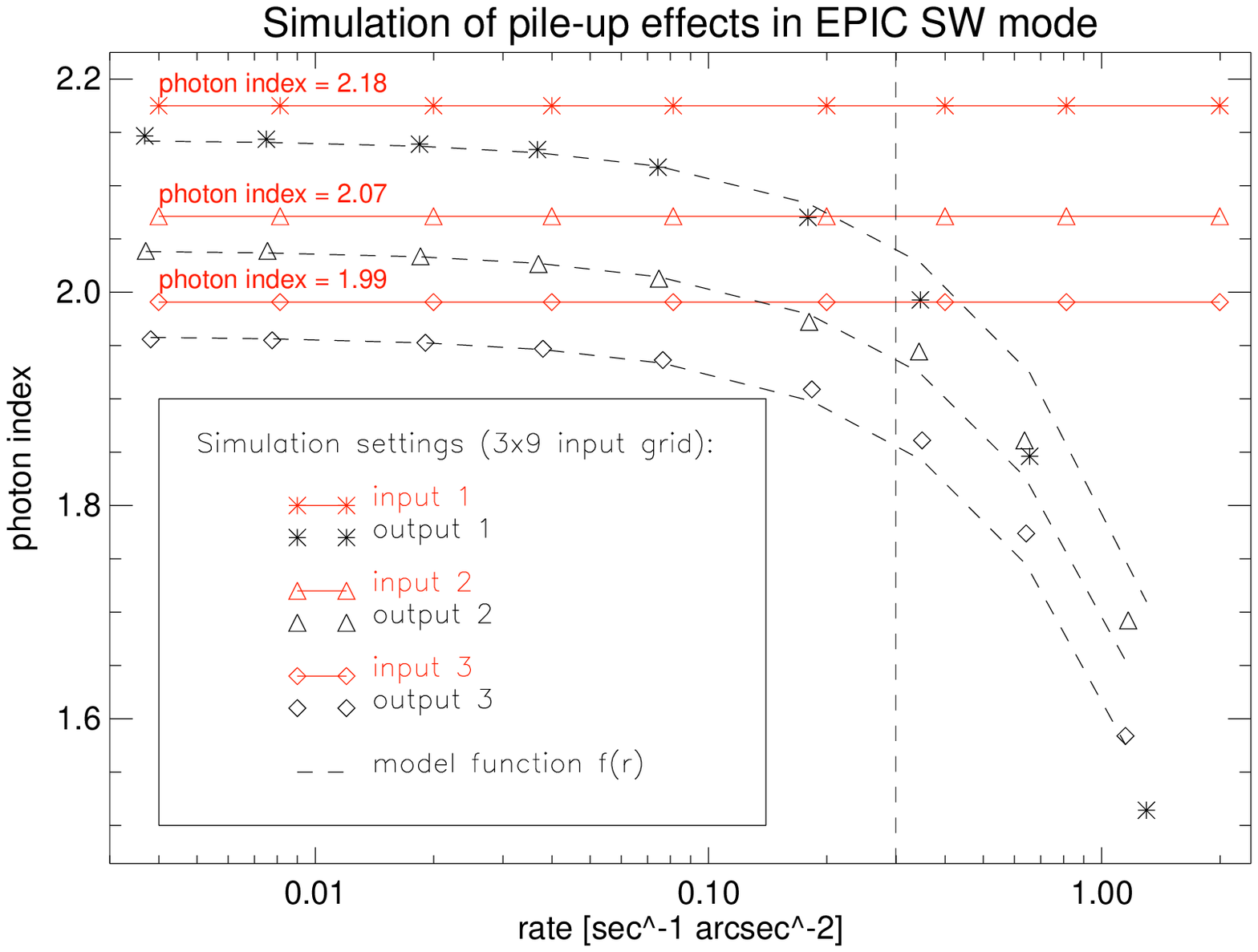}}
\caption{Result of Monte Carlo simulations to estimate the effect of
  pile-up to the measured spectra depending on photon index and count
  rate.}
\label{sim}
\end{figure}

Fig.~\ref{sim} shows the result of our simulations and indicates the shape of the correction 
function for different initial settings.

The corrections for the photon index have been applied to the Small Window mode data. The
derived correction function (\ref{eq:corr}) is in very good accordance to our independent
observational quantification of pile-up effects for the one-dimensionally averaged Small
Window data (Fig.~\ref{sw_bu_pi}, right). Correction of the two dimensional spectroscopy of
the Crab yields a final photon index of $\Gamma=1.96$ (without correction: $\Gamma=1.6$) for
the innermost extraction region around the Crab Pulsar.

\begin{figure}
\resizebox{\hsize}{!}{\includegraphics[bb=50 380 540 740,clip=,]{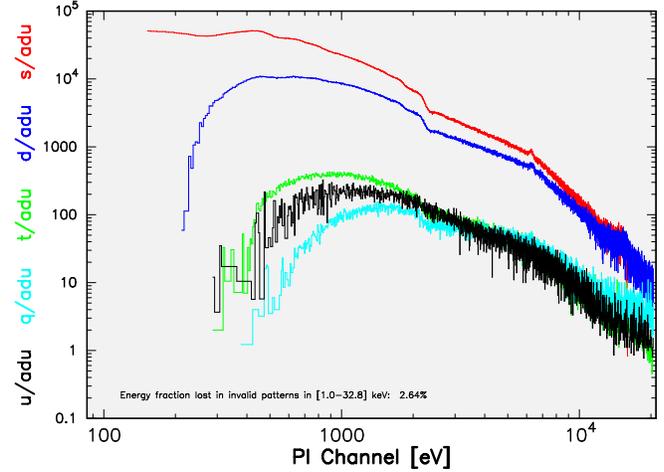}}
\caption{Distribution of patterns for a merged EPIC-pn Small Window mode observation 
of 620 ks public data. Internal background has been subtracted using
125 ks merged Closed-Filter observations. Red: singles, blue: doubles, green: triples, 
light blue: quadruples, black: invalid patterns. }
\label{epat}
\end{figure}

Note that for that analysis triple and quadruple events can be neglected in the case of the EPIC-pn camera
since they only account for less than 4\% of all events, see also Fig.~\ref{epat}.

\section{Conclusion}\label{sec:discuss}


Our work provides phase resolved spectroscopy of the modulated
flux of the Crab Pulsar in combination with spatially dependent
spectral results of the pulsar and the nebula. We measure a temporal
hardening of the spectral index in the inter peak region by 0.4.
This confirms the results achieved by \textsl{Beppo-SAX}
\citep{2000A&A...361..695M} and RXTE \citep{1997ApJ...491..808P}
measurements. In addition, we give some insight into the spatial
behaviour of the nebula both from one-dimensional EPIC-pn Burst mode
and two-dimensional EPIC-pn Small Window mode data. We show that the pulsar
region of the Crab Nebula has by far the hardest spectrum getting
softer at larger distance from the pulsar. These results are in line with
the analysis of \textsl{XMM-Newton} EPIC-MOS data \citep{willing01} and
Chandra \citep{2004ApJ...609..186M}.

For the first time we can verify a simulated pile-up correction for
\textsl{XMM-Newton} imaging mode data with real data from the one
dimensionally resolved Burst mode.

We like to stress that this paper describes in detail how \textsl{XMM-Newton}
EPIC-pn data from its special Burst mode need to be analysed
addressing both the excellent prospectives and the strength of that
mode, but also the caveats that are crucial for an adequate treatment
of the data and their exploitation.


\begin{acknowledgements}

  The \textsl{XMM-Newton} project is an ESA Science Mission with
  instruments and contributions directly funded by ESA Member States
  and the USA (NASA). The German contribution of the
  \textsl{XMM-Newton} project is supported by the Bundesministerium
  f\"{u}r Bildung und Forschung/Deutsches Zentrum f\"{u}r Luft- und
  Raumfahrt. The UK involvement is funded by the Particle Physics and 
  Astronomy Research Council (PPARC).

\end{acknowledgements}


\end{document}